# Virtual and Real Data Populated Intersection Visualization and Testing Tool for V2X Application Development

Sukru Yaren Gelbal, Mustafa Ridvan Cantas, Bilin Aksun Guvenc, and Levent Guvenc
Automated Driving Lab, Ohio State University

## Abstract

Connected Vehicle (CV) technologies have been progressing rapidly in the US. The capability afforded by Vehicle-to-Vehicle (V2V) communication improves situational awareness and provides advantages for many of the traffic problems caused by reduced visibility or No-Line-of-Sight situations, being useful for both autonomous and non-autonomous driving. Additionally, with the traffic light Signal Phase and Timing (SPaT) and Map Data (MAP) information and other advisory information provided with Vehicle-to-Infrastructure (V2I) communication, outcomes which benefit the driver in the long run, such as reducing fuel consumption with speed regulation or decreasing traffic congestion through optimal speed advisories, providing red light violation warning messages and intersection motion assist messages for collision-free intersection maneuvering are all made possible. However, developing applications to obtain these benefits requires an intensive development process within a lengthy testing period. Understanding the intersection better is a large part of this development process. Being able to see what information is broadcasted and how this information translates into the real world would both benefit the development of these highly useful applications and also ensure faster evaluation, when presented visually, using an easy to use and interactive tool. Moreover, recordings of this broadcasted information can be modified and used for repeated testing. Modification of the data makes it flexible and allows us to use it for a variety of testing scenarios at a virtually populated intersection. Based on this premise, this paper presents and demonstrates visualization tools to project SPaT, MAP and Basic Safety Message (BSM) information into easy to read real-world based graphs. Also, it provides information about the modification of the real-world data to allow creation of a virtually populated intersection, along with the capability to also inject virtual vehicles at this intersection.

## Introduction

While the number of the vehicles on the roads [1] and their automated driving capabilities [2], [3], [4], [5], [6], [7] increase, concerns for their safety [8], [9], [10], [11], [12] traffic congestion and fuel economy [13] increase. In order to reduce these problems, V2X technology is rapidly growing and many applications are being developed to provide better overall travel experience. With the utilization of V2X information, improvements can be significant for a wide variety of aspects. For example, information can be leveraged to detect traffic rule violations [14, 15] or predict future threats to warn the driver and provide safer experience for all road users. Traffic light timings can be adaptively changed in cooperation with the vehicles on the road [16, 17], to increase travel efficiency and reduce the traffic congestion. Traffic light state and remaining time information can be utilized to provide an optimized speed advisory to the driver [18, 19, 20] in order to achieve a more uniform driving experience and better fuel economy.

While the design and processing capability of these applications are important, adequate representation of the information is also equally important. Representing the information with correct visualization is highly beneficial. One objective of the visualization for V2X information is visualization for better driver awareness. Within this objective, in order to convey the information obtained through the applications as robustly and as fast as possible to the driver, design of the Human Machine Interface (HMI) is critical. Reference [21] investigates many different approaches in the literature about HMI designs applied to different CV applications and shares important conclusions for achieving the objective of better driver awareness of the current traffic situation.

Another objective of the visualization for V2X information is to assist the development of these CV applications. With visualization of these messages, one can provide an easier understanding of the meaning behind the messages, which increases the effectiveness for the development process of the applications and ensures faster evaluation of V2X messages broadcasted by CVs and infrastructure equipment. There are several tools available and developed by the United States Department of Transportation (USDOT) as web applications [22]. These tools can be used to create encoded MAP, SPaT messages from a visual interface and validate encoded messages of different types. As seen from the literature and by the efforts from USDOT, the demand for such utility tools is increasing.

Hence, this paper aims to contribute with tools developed by the authors to translate SPaT, MAP and BSM messages into their visual representations based on real world graphs. These tools can be used as a utility in the process of developing CV applications or can be used for evaluation of the V2X data broadcasted by the V2X users. Data from real world intersections were collected with a connected automated vehicle and post-processed in order to demonstrate the visualization tool developed and presented in this paper. Moreover, the paper introduces how collected data can be used to create virtually populated intersections and provides a proof of concept experiment using this method. These virtually populated intersections can be used for CV application testing with realistic data.

The remainder of the paper is organized as follows. The Data section discusses the data used for the visualization, provides information about the intersections this data belongs to and provides brief information about the vehicle used for data collection. This is followed by the Visualization section which talks about the workflow from data to graphs, then demonstrates visualization of the data on selected intersections in two sub-sections which are visualization of intersection geometry and signal phase, and visualization of road users. Afterwards, a method to use this data to populate an intersection virtually is discussed along with a proof of concept experiment, in the Virtually Populating Intersections section. The paper ends with the Conclusion and Future Work section which provides a summary of the work that has been reported within this study and discusses possible improvements and application fields that can be realized in the future.

## Data

There are several RSU implementations around the Columbus, Marysville and Dublin areas in Ohio. Because of the public availability and realistic nature of the data, two of these intersections were selected to collect the data to demonstrate the visualization tools discussed in this paper. The base data used in this study contains SAE 2735 [23] SPaT, MAP and BSM messages broadcasted by these two intersections in Packet Capture (PCAP) format. The vehicle used for data collection and these intersections are discussed in this section along with brief information about the intersection geometry and broadcasted messages.

### First Intersection

This intersection is located at Hilltop, Columbus, near the Ohio Public Safety Department. It is a straight road with a pedestrian crosswalk. The road has three lanes on each side of the intersection, two lanes approaching and one lane departing. A top-down view of this intersection is shown in Figure 1. The intersection reference point is shown with a red marker.

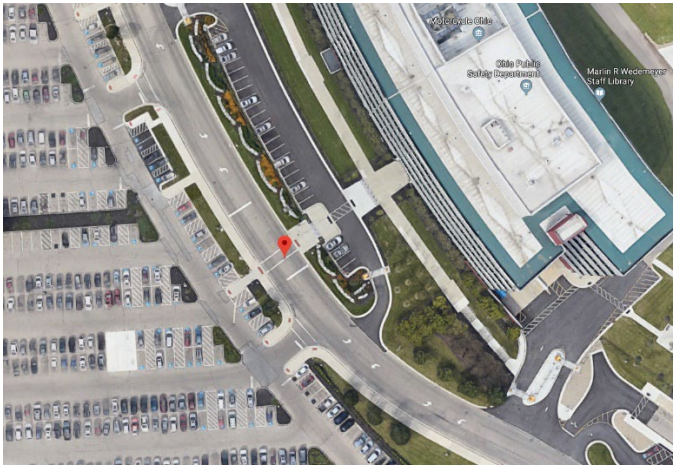

Figure 1. Top-down view of the first intersection selected.

This intersection broadcasts SPaT and MAP messages. One special feature of this intersection is that the traffic light state is always green unless a pedestrian pushes the designated button to cross the road. Therefore, in the SPaT message, if the signal state is green, remaining time information goes up one second right after it reaches zero and repeats, as long as the pedestrian button is not pressed. This intersection was selected to visualize the intersection geometry obtained through the MAP message and to visualize the current signal phase and remaining time obtained through the SPaT message.

### Second Intersection

This intersection is located at the intersection of East 5th Street & North Main Street, Marysville, Ohio. It is a four-way intersection with pedestrian crosswalks being present for each of the four sides. Each side has a three lane road with two lanes approaching and one lane departing. A top-down view of this intersection is shown in Figure 2. The intersection reference point is shown with a red marker.

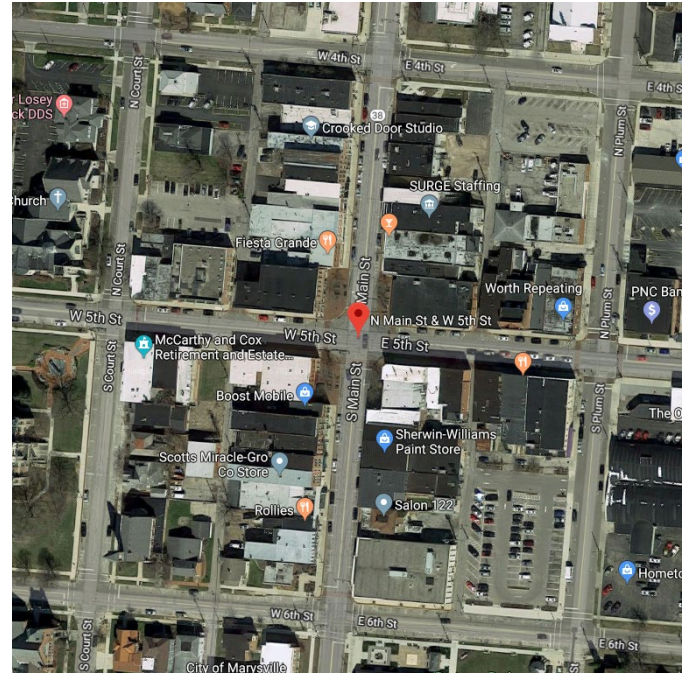

Figure 2. Top-down view of the second intersection selected.

This intersection has a special case. It is a smart intersection with cameras equipped and looking down at all four directions, to detect and track vehicles and pedestrians [24]. It classifies these road users within several classes, which are normal vehicle, pedestrian, motorcycle and emergency vehicle. Along with the classification, it calculates positions, headings and speeds of these road users, then publishes BSM messages about the ones approaching towards the intersection, through the implemented RSU. This creates a very unique environment as if all of the road users were equipped with V2V equipment and broadcasting messages about themselves. As a result, this smart intersection enables testing of V2V based applications very realistically. The main reason that this intersection was selected is the availability of a high number of BSM messages. These BSM messages were used to obtain information about the road users and visualize them.

### Data Collection

For the first intersection, data was recorded by a DENSO 5900A On Board Unit (OBU), fixed in the trunk of the automated Ford Fusion Hybrid connected and autonomous research vehicle of the Automated Driving Lab of the Ohio State University. Although the vehicle is automated [25] and these autonomous driving features can be combined with CV applications in the future, for the purpose of this study, the vehicle was manually driven on the road to collect data from the intersection. All the data was listened to through channel

172 and directly recorded as a PCAP file as it was being published from the intersection. A picture of the vehicle along with the OBU device in the trunk and the placement of the antenna for the OBU are shown in Figure 3.

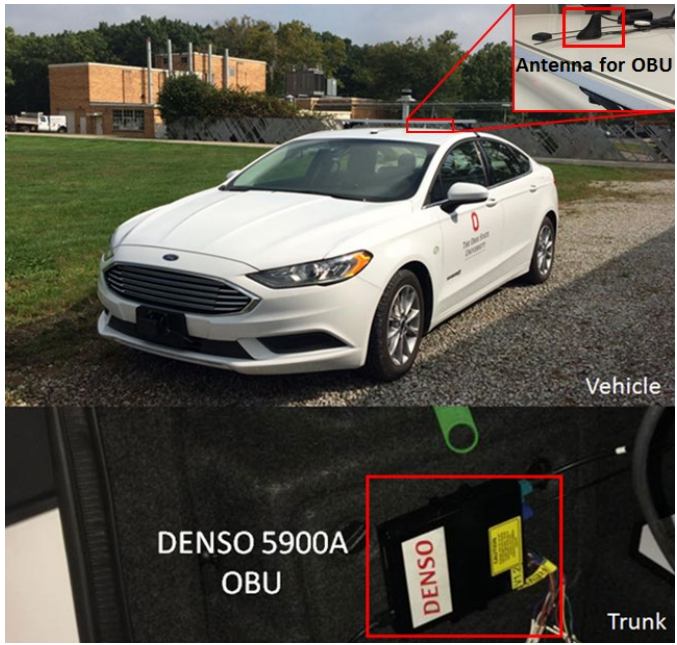

Figure 3. Experimental vehicle and OBU device used for data collection.

For the second intersection, initial testing and data collection were carried out using this experimental vehicle as well. After the testing and initial phases were finished, the data used for demonstration of the visualization in this study was recorded and provided to us by the City of Marysville, again in PCAP format.

## Visualization

In this section, the workflow used is discussed and the visualization algorithms are demonstrated on the intersections discussed in the previous section.

### Data Flow

Data broadcasted by the RSU is recorded directly using the OBU inside the vehicle. Since the recorded data is in PCAP format, it needs to be processed in several steps before it is visualized. These steps are shown schematically in the diagram of Figure 4.

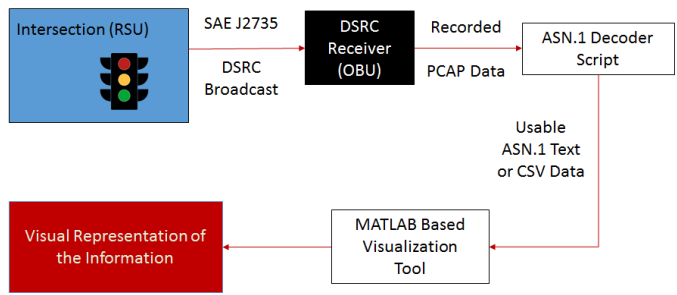

Figure 4. Data flow for visualization.

As seen in the figure above, after recording, the data needs to be decoded. In order to decode data, depending on the output format necessary, a Python script and a C program that uses the ASN1 compiler [26] library were prepared. These scripts decode the data according to the Unaligned-Packet Encoding Rules (UPER) rule, which is the standard decoding rule for SAE J2735 2016-03. After the data is decoded, it is written to a separate file in ASN.1 readable text format for MAP message or as Comma Separated Value (CSV) format for the BSM message. After the data is converted into one of these formats, it is imported into MATLAB to be used by the visualization tools. By using this data, the visualization tools create graphs based on the visual representation of the data from several aspects. There are currently two different visualization tools which are explained in the following sections.

### Visualization of Intersection Geometry and Signal Phase

The purpose of this tool is to visualize the information about the lane nodes and intersection reference point through MAP message and signal phase and remaining time through the SPaT message. All of these messages were broadcasted by the RSU implemented at the intersection and collected by the OBU in our experimental vehicle.

The tool is designed to draw lane nodes on top of a top-down view image of the intersection. These lanes nodes are several points determined to reside on each lane. Node sets are created for each lane and nodes belonging to the same lane are drawn with the same color. The intersection reference point, which is the global coordinate center that these lane nodes are based upon (which most of the time indicates the center of the intersection), is shown as a plus sign (+). After drawing the information about the intersection geometry, the visualization tool draws traffic lights for each phase group at the intersection and displays the traffic light state with red, green, yellow colored circles depending on the phase and displays remaining time at the bottom as a text. An image generated by this tool is shown in Figure 5.

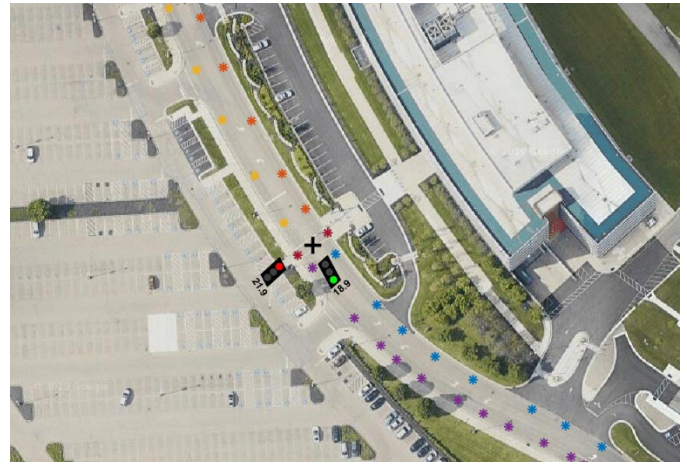

Figure 5. Generated image by visualization of intersection geometry and signal phase.

The intersection used for demonstration of the tool is the first intersection discussed in the Data section in the paper. As mentioned above, each lane has a set of nodes that are drawn with color. Same color means those nodes belong to same lane. In the picture above, five different lanes can be seen with yellow, orange, magenta, blue colored vehicle lanes and dark red colored crosswalk. The intersection reference point is marked with a plus sign (+). Signal phase groups can also be seen as traffic lights drawn with colored circles. Traffic light seen on the left belongs to the crosswalk crossing and traffic light on the right belongs to vehicles crossing the intersection. Traffic light states (signal phase) are shown with green and red colored circles, with current remaining time in seconds being shown at the bottom, as text.

Since the tool is able to repeat this signal phase and timing drawing process for every frame, the data can be used to create a continuous visualization. The visualization tool software was later modified to repeat the process for every certain time step, effectively allowing for creation of a video from the data. This feature was used to create a video, showing the remaining time constantly changing according to the data, as well as the intersection states. Then, this video was recorded and uploaded to a file service [15] and also uploaded along with the paper. In the video, traffic light state and remaining time can be seen to be changing depending on the data.

### *Visualization of Road Users*

The purpose of this tool is to visualize information about the road users in the vicinity, obtained through BSM messages. In our case, these messages are provided by an RSU implemented on the smart intersection that was discussed in the Data section. In the future, when V2X equipment becomes more common in cars, this information will be available at any place where connected road users exist. In Smart Columbus, this data will be available near real-time through a websocket protocol connection to the Smart Columbus Operating System, for example.

The tool is currently designed to visualize the location, heading and ID of the road users. Since the location information is available in the BSM message in latitude and longitude, heading in degrees and ID in hex format, the tool is able to visualize these. The tool reads all the locations and headings of the road users at any given time, determines the color of the cursor using the ID in the corresponding BSM, then draws rectangles on top of a top-down view image of the intersection with the correct heading for vehicles. In case there is a pedestrian, the tool draws a red circle at the position of the pedestrian. The tool repeats this process for all the road users present for the given timeframe. An image generated by this tool is shown in Figure 6.

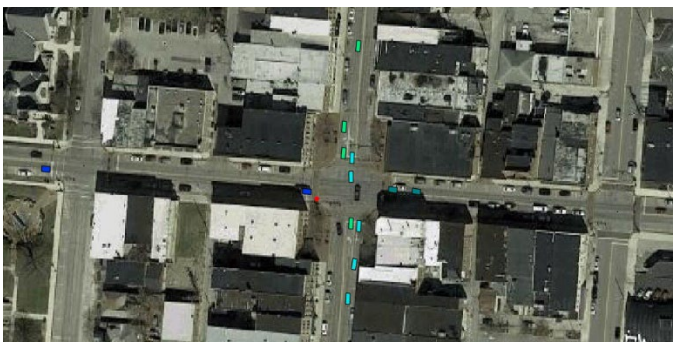

Figure 6. Generated image by visualization of road users.

The intersection used here is the second intersection discussed in the Data section. Rectangles represent the vehicles where each color is generated from their ID. It is seen that there are multiple vehicles approaching the intersection. It is important to note that the data contains approaching vehicles only, because the detection system is implemented as such. There is also a pedestrian crossing the street that is noticeable at the left side of the intersection in the middle of Figure 6, represented by a red circle.

Similar to the previous tool, this visualization tool can also repeat the drawing process continuously, therefore allowing creation of a video from the data. This feature was used to create a video for this visualization as well. Then, the video was recorded and uploaded to a file service [16] and also uploaded along with the paper. In the video, the road users can be seen moving, changing lanes and crossing the intersection.

## Virtually Populating Intersections

Along with the visualization for informing the driver or for enhancing the analysis process, the collected data during this study can also be highly beneficial for CV application testing. This section discusses a method for virtually populating intersections with real traffic and demonstrates a proof of concept testing approach for a CV application at this virtually populated intersection.

### *Moving the Data to Another Intersection*

Recorded data is saved as PCAP files. These files can be replayed using a DSRC device, in our case an additional OBU, which results in a stream of V2X messages that can be captured and processed by other listeners around the area. These messages can be anything captured in the PCAP file. For the purpose of populating the intersection with virtual vehicles, we isolate BSM messages from this file. By isolating these messages and creating a separate file, replayed DSRC messages will only consist of BSM messages, which will serve the purpose of populating intersection with virtual vehicles while preventing any intervention with the infrastructure based messages such as SPaT or MAP.

Another key point for this implementation is to have positions, headings and elevations of the virtual vehicles matched with the intersection that is virtually populated. In order to do this, a Python script was coded to read, decode the PCAP file, add an offset to the locations, adjust headings of vehicles, encode back and create a modified file with same format. This process can be done for any intersection given the information of latitude, longitude, heading and elevation offsets between center of the corresponding intersection and the intersection that the data is collected from. Obviously, a complete match of lanes shouldn't be expected since intersections are highly variable in terms of geometry, unless data is collected from a specific intersection that exactly matches the geometry of the intersection that is planned to be virtually populated. If an exact match is essential for the purpose of the test, messages in PCAP file can be manipulated in such way to artificially move the vehicles one by one to match the lane positions. For the scope of this study, exact lane match was not required, therefore an offset was applied to all the vehicles in the data to match the vehicles to the desired intersection.

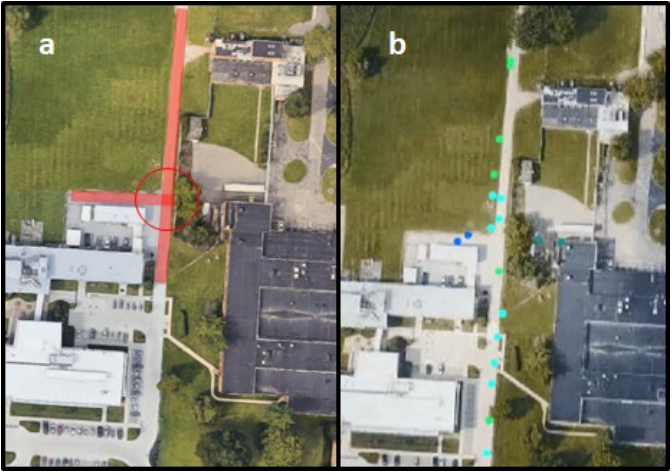

Figure 7. (a) Selected intersection to be populated. (b) Virtual vehicles shown on the intersection.

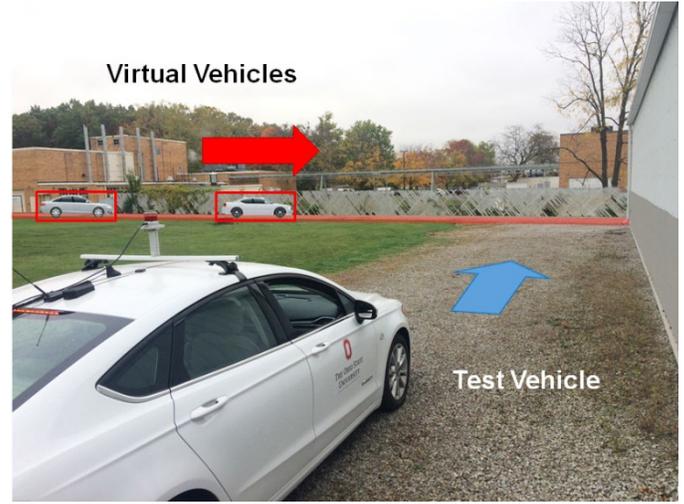

Figure 8. Experiment conducted with the test vehicle and virtual vehicles.

The selected intersection to be populated is a dirt road near our lab building, shown in Figure 7a and is marked with red lines. Source data comes from the intersection located at Marysville, which was also discussed above, under the Data section. At the end of the modification process, virtual vehicles can be seen as visualized on top of the intersection with colored circles in Figure 7b, approaching towards the intersection. They are colored in various tones of blue and green according to their ID value.

As a result of this process, we have virtual vehicles approaching towards the intersection and interacting with each other. Moreover, this data is highly realistic since it is collected from real vehicles driving at another intersection. Therefore, it can be highly beneficial for CV application testing. The next section discusses a proof of concept experiment for testing a CV application at this virtually populated intersection.

### *Testing CV Applications*

A proof of concept testing of an Intersection Movement Assist (IMA) CV application was conducted at this intersection to see if the vehicle is able to process the information and issue the warning correctly. For the purpose of this test, the data is further isolated such that it only contains the vehicles approaching from the North side. You can see a depiction of the experiment in Figure 8, where the test vehicle approaches towards the intersection at the same time as the virtual vehicles illustrated in the image.

A simplified IMA algorithm was implemented in the OBU to conduct a proof of concept experiment using this environment. This algorithm listens to BSM messages, processes them, informs the driver about the vehicles approaching and in case of any predicted collision, issues an IMA warning for the driver to slow down or stop to prevent the collision. Warning calculation is done by first calculating time to intersection for host and remote vehicles.

$$t_{remote} = \frac{d_{remote}}{v_{remote}} \qquad (1)$$

$$t_{host} = \frac{d_{host}}{v_{host}} \qquad (2)$$

where $t$ represents time to intersection, $d$ represents the distance and $v$ represents the speeds for remote and host vehicles. After times $t_{remote}$ and $t_{host}$ are calculated, if the times are close to each other within a margin of safety, an IMA warning is issued.

$$t_{remote} - \frac{t_{safety}}{2} < t_{host} < t_{remote} + \frac{t_{safety}}{2} \qquad (3)$$

$t_{safety}$ is the safety margin mentioned above and was assigned as 3 seconds in the current experiment. If the condition of Equation 3 is satisfied, an IMA warning is issued.

The interface for the application is a terminal based interface, which can be seen in Figure 9. Two snapshots were taken from the interface at two different times. Figure 9a shows the information shown to the driver about approaching vehicle locations, speeds, headings and distances. Figure 9b was taken when there is a collision risk with one of the approaching vehicles and an IMA warning was issued. Warning can be seen with capital letters, as well as the reasonable speed and distance of the remote vehicle approaching, for the warning to be issued.

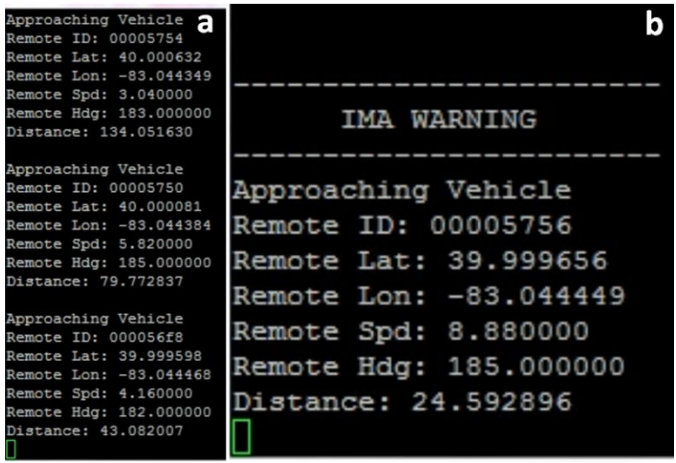

Figure 9. (a) Driver is informed about approaching vehicles. (b) IMA warning is issued if collision is predicted.

As it can be seen from Figure 9, while the data is replayed, virtual vehicles were captured by the OBU, correctly processed and shown to the driver. IMA warning was also correctly issued when there was a collision predicted.

## Conclusions and Future Work

CV technology is a rapidly growing and developing field which is currently receiving a high amount of attention from both researchers and automotive companies. Applications developed by leveraging information obtained from V2X communication can be highly beneficial for both automated and non-automated driving. These benefits can cover many different aspects such as safety, shorter travel time and better fuel economy. Based on this premise, this paper presented utility tools for projecting the information from V2X messages such as SPaT, MAP and BSM, onto real world based easily readable graphs. This allows understanding the V2X information better, quicker and as a result, makes it easier to troubleshoot or evaluate CV applications and V2X implementations. Moreover, the paper also discussed a method where collected data can be used to virtually populate intersections, which allows more realistic and safe testing of CV apps. Since the data source is the real world intersections, testing will be highly realistic.

One of the first steps as future work would be to wrap up the tools presented in this study for public use and upload them to a public website, such as GitHub. There are also several improvements which can be implemented in the future. Current design of the data flow is intended to only allow offline data processing and visualization. In the future, data flow can be modified to achieve real-time online visualization of the V2X data, which would increase the benefits of these tools considerably. These tools can also be combined in the future to show all of the information, even from multiple intersections, on the same graph, with an improved design that will not cause confusion. Finally, as an addition for the virtually populated intersections, the repeatable nature of the virtual vehicle behavior can be utilized for CV application improvements that require repetition and comparison, such as parameter tuning.

The playback of real CV data and use of virtual vehicles that was presented in this paper can be used in safe experimental testing and validation of many vehicle control, robust control, ADAS, connected and automated driving methods in a realistic closed testing environment while being immersed in realistic traffic and infrastructure emulation [27-56].

13. A. Boyali and L. Guvenc, "Real-Time Controller Design for a Parallel Hybrid Electric Vehicle Using Neuro-Dynamic Programming Method," in *IEEE Systems, Man and Cybernetics*, İstanbul, October 10-13, pp. 4318-4324, 2010.
14. D. Anushya, "Vehicle Monitoring for Traffic Violation Using V2I Communication," in *Second International Conference on Intelligent Computing and Control Systems (ICICCS)*, Madurai, India, 2018.
15. S. Y. Gelbal, M. R. Cantas, B. Aksun-Guvenc, L. Guvenc, G. Surnilla, H. Zhang, M. Shulman, A. Katriniok and J. Parikh, "Hardware-in-the-Loop and Road Testing of RLVW and GLOSA Connected Vehicle Applications," in *SAE World Congress Experience*, (Currently Under Review), 2020.
16. J. Li, Y. Zhang and Y. Chen, "A Self-Adaptive Traffic Light Control System Based on Speed of Vehicles," in *IEEE International Conference on Software Quality, Reliability and Security Companion (QRS-C)*, Vienna, Austria, 2016.
17. B. Xu, X. J. Ban, Y. Bian, J. Wang and K. Li, "V2I based cooperation between traffic signal and approaching automated vehicles," in *IEEE Intelligent Vehicles Symposium (IV)*, Los Angeles, CA, USA, 2017.
18. A. Stevanovic, J. Stevanovic and C. Kergaye, "Green Light Optimized Speed Advisory Systems," *Transportation Research Record: Journal of the Transportation Research Board,* vol. 2390, pp. 53-59, 2013.
19. M. R. Cantas, O. Kavas, S. Tamilarasan, S. Y. Gelbal and L. Guvenc, "Use of Hardware in the Loop (HIL) Simulation for Developing Connected Autonomous Vehicle (CAV) Applications," in *SAE World Congress Experience*, 2019.
20. Gelbal, S.Y., Cantas, M.R, Tamilarasan, S., Guvenc, L., Aksun-Guvenc, B., "A Connected and Autonomous Vehicle Hardware-in-the-Loop Simulator for Developing Automated Driving Algorithms," in *IEEE Systems, Man and Cybernetics Conference*, Banff, Canada, 2017.
21. C. Olaverri-Monreal and T. Jizba, "Human Factors in the Design of Human–Machine Interaction: An Overview Emphasizing V2X Communication," *IEEE Transactions on Intelligent Vehicles,* vol. 1, pp. 302-313, 2016.
22. "Connected Vehicle Tools," USDOT, [Online]. Available: https://webapp2.connectedvcs.com/. [Accessed November 2019].
23. SAE, "J2735 Dedicated Short Range Communications (DSRC) Message Set Dictionary," 2016.
24. "Honda Demonstrates New "Smart Intersection" Technology," [Online]. Available: https://csr.honda.com/2018/10/04/honda-demonstrates-new-smart-intersection-technology/. [Accessed November 2019].
25. S. Y. Gelbal, B. Aksun-Guvenc and L. Guvenc, "SmartShuttle: A Unified, Scalable and Replicable Approach to Connected and Automated Driving in a SmartCity," in *Second International Workshop on Science of Smart City Operations and Platforms Engineering (SCOPE)*, Pittsburg, PA, 2017.
26. L. Walkin, "ASN1 C Compiler," [Online]. Available: https://github.com/vlm/asn1c. [Accessed November 2019].
27. H. Wang, S. Gelbal and L. Guvenc, ""Multi-Objective Digital PID Controller Design in Parameter S Robust PID Steering Control pace and its Application to Automated Path Following," *IEEE Access,* vol. 9, pp. 46874-46885, 2021.
28. B. Demirel and L. Guvenc, "Parameter Space Design of Repetitive Controllers for Satisfying a Mixed Sensitivity Performance Requirement," *IEEE Transactions on Automatic Control,* vol. 55, pp. 1893-1899, 2010.
29. B. Aksun-Guvenc and L. Guvenc, "Robust Steer-by-wire Control based on the Model Regulator," in *IEEE Conference on Control Applications*, 2002.
30. B. Orun, S. Necipoglu, C. Basdogan and L. Guvenc, "State Feedback Control for Adjusting the Dynamic Behavior of a Piezo-actuated Bimorph AFM Probe," *Review of Scientific Instruments,* vol. 80, no. 6, 2009.
31. L. Guvenc and K. Srinivasan, "Friction Compensation and Evaluation for a Force Control Application," *Journal of Mechanical Systems and Signal Processing,* vol. 8, no. 6, pp. 623-638.
32. M. Emekli and B. Aksun-Guvenc, "Explicit MIMO Model Predictive Boost Pressure Control of a Two-Stage Turbocharged Diesel Engine," IEEE Transactions on Control Systems Technology, vol. 25, no. 2, pp. 521-534, 2016.
33. Aksun-Guvenc, B., Guvenc, L., 2001, "Robustness of Disturbance Observers in the Presence of Structured Real Parametric Uncertainty," Proceedings of the 2001 American Control Conference, June, Arlington, pp. 4222-4227.
34. Guvenc, L., Ackermann, J., 2001, "Links Between the Parameter Space and Frequency Domain Methods of Robust Control," International Journal of Robust and Nonlinear Control, Special Issue on Robustness Analysis and Design for Systems with Real Parametric Uncertainties, Vol. 11, no. 15, pp. 1435-1453.
35. Demirel, B., Guvenc, L., 2010, "Parameter Space Design of Repetitive Controllers for Satisfying a Mixed Sensitivity Performance Requirement," IEEE Transactions on Automatic Control, Vol. 55, No. 8, pp. 1893-1899.
36. Ding, Y., Zhuang, W., Wang, L., Liu, J., Guvenc, L., Li, Z., 2020, "Safe and Optimal Lane Change Path Planning for Automated Driving," IMECHE Part D Passenger Vehicles, Vol. 235, No. 4, pp. 1070-1083, doi.org/10.1177/0954407020913735.
37. T. Hacibekir, S. Karaman, E. Kural, E. S. Ozturk, M. Demirci and B. Aksun Guvenc, "Adaptive headlight system design using hardware-in-the-loop simulation,"2006 IEEE Conference on Computer Aided Control System Design, 2006 IEEE International Conference on Control Applications, 2006 IEEE International Symposium on Intelligent Control, Munich, Germany, 2006, pp. 915-920, doi: 10.1109/CACSD-CCA-ISIC.2006.4776767.
38. Guvenc, L., Aksun-Guvenc, B., Emirler, M.T. (2016) "Connected and Autonomous Vehicles," Chapter 35 in Internet of Things/Cyber-Physical Systems/Data Analytics Handbook, Editor: H. Geng, Wiley.
39. Emirler, M.T.; Guvenc, L.; Guvenc, B.A. Design and Evaluation of Robust Cooperative Adaptive Cruise Control Systems in Parameter Space. International Journal of Automotive
Page 7 of 8

## Contact Information

Sukru Yaren Gelbal

The Ohio State University

Electrical and Computer Engineering Department

gelbal.1@osu.edu

1320 Kinnear Road, Columbus, OH, 43212

## Acknowledgments

The authors would like to thank Mike Andrako from the City of Marysville in Ohio for providing V2X data used for visualization of road users. The donation of the OBU device which was used for data collection in this study by the DENSO Corporation is gratefully acknowledged.

## Abbreviations

| | |
|---|---|
| **CV** | Connected Vehicle |
| **V2V** | Vehicle to Vehicle |
| **SPaT** | Signal Phase and Timing |
| **MAP** | Map Data |
| **V2I** | Vehicle to Infrastructure |
| **BSM** | Basic Safety Message |
| **HMI** | Human Machine Interface |
| **USDOT** | United States Department of Transportation |
| **PCAP** | Packet Capture |
| **OBU** | On Board unit |
| **UPER** | Unaligned-Packet Encoding Rules |
| **CSV** | Comma Separated Value |
| **IMA** | Intersection Movement Assist |